\documentclass{jetpl}
\twocolumn
\lat

\title{Distorted vortex lattice in a tetrahedral superconductor} 
\rtitle{Distorted vortex lattice in a tetrahedral superconductor} 
\sodtitle{Distorted vortex lattice in a tetrahedral superconductor} 

\author{V.\,H. Dao\/\thanks{dao@drfmc.ceng.cea.fr}  and 
M.\,E. Zhitomirsky}
\rauthor{V.\,H. Dao, M.\,E. Zhitomirsky}
\sodauthor{V.\,H. Dao, M.\,E. Zhitomirsky}

\address{
Commissariat \`a l'Energie Atomique, DSM/DRFMC/SPSMS, 38054 Grenoble,
France}
\dates{13 January 2006}{*}

\abstract{
Equilibrium shape and orientation of vortex lattice
are studied for an $s$-wave tetrahedral superconductor in the vicinity
of the upper critical field. The phase diagram,
which includes transitions between 
rhombic and rectangular lattices, is constructed in the 
parameter space of the Ginzburg-Landau functional.
The developed theory is applied to the heavy-fermion superconductor
PrOs$_4$Sb$_{12}$. 
In a wide range of parameters the shape of the vortex lattice
is only weakly dependent on temperature.
The neutron scattering measurements of the vortex lattice in
PrOs$_4$Sb$_{12}$ can be explained by a peculiarities of the
tetrahedral symmetry group and are further supported
by analysis of the appropriate band structure calculations.
}

\begin{document}

\maketitle

Hexagonal vortex lattice predicted for ideal isotropic 
superconductors \cite{abrikosov,saint-james} is perturbed in real materials by
crystalline anisotropy. Anisotropic nonlocal corrections within 
the Ginzburg-Landau theory or in the London approximation are determined 
by the Fermi surface geometry \cite{takanaka}.
Their effect may result in a 
hexagonal-to-square vortex lattice transition observed in the past
in the superconducting borocarbides
\cite{yaron,dewilde,park,klironomos}.
Anisotropy of the Cooper pairs wave function 
also gives rise to vortex lattice distortions.  
The line nodes of the $d_{x^2-y^2}$ superconducting gap in the high-$T_c$
cuprates favor, for example, a square vortex lattice \cite{won, affleck}.
The flux line lattice (FLL) geometry provides, thus, a combined insight into the
Fermi surface and the order parameter anisotropy.

Filled skutterudite compound PrOs$_4$Sb$_{12}$ has attracted 
significant attention in the past few years as a first example of
Pr-based heavy fermion superconductor with $T_c=1.85$~K 
\cite{bauer,maple,vollmer,measson,seyfarth}.
At present, controversy remains about the symmetry and the structure
of the superconducting gap in PrOs$_4$Sb$_{12}$. The low-temperature 
scanning tunneling spectroscopy experiments
by Suderow {\it et al\/}.~\cite{suderow} clearly demonstrate  
a finite superconducting gap over a large part of the Fermi surface.
This finding is independently confirmed by an exponential low-temperature 
dependence
of the nuclear relaxation rate $1/T_1$ \cite{kotegawa},
though absence of the coherence peak may point to anisotropic pairing.
Recently, small angle neutron scattering measurements have found a distorted
vortex lattice in PrOs$_4$Sb$_{12}$ at low fields and temperatures
\cite{huxley}.
The observed distortion was attributed by Huxley {\it et al\/}.\ to an
anisotropic multicomponent superconducting order parameter with point nodes.
PrOs$_4$Sb$_{12}$ has, however, a rather unusual tetrahedral
$T_h$ point group. 
This group contains three-fold rotations about cube diagonals but 
no four-fold rotations in contrast to the other cubic groups $O$ and $O_h$. 
Investigation of the shape of FLL has not been done to our knowledge
for such superconductors.
In the present work we investigate the geometry of FLL in tetrahedral
superconductors within the nonlocal Ginzburg-Landau theory for a conventional
$s$-wave order parameter. 

The Ginzburg-Landau (GL) energy density functional for an $s$-wave 
superconducting order parameter $\Psi$ derived from 
the BCS theory has the standard form: 
\begin{equation}
F_{GL} = 
\Psi^*({\bf x})\left( \alpha  + {\cal F}_{\Pi} \right) \Psi({\bf x}) +
\frac{\beta}{2} |\Psi({\bf x})|^4 + \frac{h^2}{8\pi} \ .
\label{GL}
\end{equation}
Here $\alpha =(1-T/T_c)$, $T_c$ is the transition temperature,
and $\beta = 7\zeta(3)/(4\pi T_c)^2$. The gradient terms 
\begin{equation}
{\cal F}_{\Pi} = \sum_{n=1}^{\infty}\sum_{i_1\ldots i_{2n}} 
K_{i_1\dots i_{2n}}
\; \; \Pi_{i_1} \ldots \Pi_{i_{2n}} 
\label{grad}
\end{equation}
are expanded into even powers of the operator
$\Pi_i= -i \partial_i + (2\pi/\Phi_0)A_i$,
$\Phi_0$ being the flux quantum. Expansion coefficients 
are expressed  via the Fermi surface averages of 
the components of the Fermi velocity ${\bf v}_F$ as 
\begin{equation}
K_{i\ldots j} = \frac{ (-1)^{n+1}}{(2\pi T_c)^{2n}} 
( 2-\frac{1}{2^{2n}} )
\zeta(2n+1)\langle v_{Fi} \dots v_{Fj} \rangle_{\rm FS}.
\label{Kij}
\end{equation}
The tetrahedral point symmetry $T_h$ imposes certain relations between 
the gradient term coefficients:
\begin{eqnarray}
&& K_{x^2} = K_{y^2} = K_{z^2},\  \ \ \ \ \ \ \ \ \ \ 
K_{x^4}=K_{y^4}=K_{z^4}, 
\nonumber \\
&& K_{x^2 y^2} = K_{x^2 y^2} = K_{y^2z^2},\ \ \ K_{x^6}=K_{y^6}=K_{z^6}, \\
&&  K_{x^4 y^2} = K_{y^4 z^2} = K_{z^4x^2}, \ \ \
K_{x^2 y^4} = K_{y^2 z^4} = K_{z^2 x^4}. \nonumber 
\end{eqnarray}
Note, that for the tetrahedral group $K_{x^4y^2} \neq K_{x^2y^4}$.

Let us now assume that an external magnetic field is applied
along the $\hat{z}$-axis. Considering only solutions, which are uniform
along the field direction ($\Pi_z\Psi\equiv 0$), the gradient
terms are simplified to
\begin{eqnarray}
{\cal F}_2 & = & K_{x^2} \left( \Pi_x^2 + \Pi_y^2 \right)\ , \nonumber \\
{\cal F}_4 & = & K_{x^4} \left( \Pi_x^4 + \Pi_y^4 \right) + K_{x^2 y^2} \{
\Pi_x^2 \Pi_y^2 \} \ , \label{F246} \\
{\cal F}_6 & = & K_{x^6} \left( \Pi_x^6 + \Pi_y^6 \right) + K_{x^4 y^2} \{
\Pi_x^4 \Pi_y^2 \} + K_{x^2 y^4} \{ \Pi_x^2 \Pi_y^4 \}. \nonumber
\end{eqnarray}
In the above expressions $\{\cdots\}$ denotes a sum over all possible
permutations of the gradient operators $\Pi_i$ and $\Pi_j$.

The upper critical field $H_{c2}$ is determined by the linearized GL equation,
which can be conveniently written in terms of the 
Landau level operators $\hat{a}$ and $\hat{a}^\dagger$:
$\hat{a}=\sqrt{\Phi_0/4\pi H}(\Pi_x-i\Pi_y)$. 
Rotation by angle $\varphi$
about $\hat{z}$-axis transforms the Landau level operators into
$e^{-i\varphi} \hat{a}$ and $e^{i\varphi} \hat{a}^{\dagger}$.
After some algebra, the gradient terms (\ref{F246}) are expressed as 
\begin{eqnarray}
{\cal F}_2 & = & h (2\hat{n}+1)\ ,  \nonumber \\
{\cal F}_4 & = & h^2 \left[ 
k_{40} ( 2 \hat{n}^2 + 2 \hat{n} +1 ) + k_{44} ( \hat{a}^4 + \hat{a}^{\dagger
4} )\right] ,  \nonumber \\
{\cal F}_6 & = & h^3 \left[ 
k_{60} \hat{n}_{60} + k_{62}(\hat{n}_{62}\hat{a}^2+
\hat{a}^{\dagger 2}\hat{n}_{62} )  \right . \nonumber  \\ 
&& \left.+ k_{64}(\hat{n}_{64}\hat{a}^4+\hat{a}^{\dagger 4}\hat{n}_{64})+k_{66}
(\hat{a}^6+\hat{a}^{\dagger 6}) \right] , 
\label{F246fin}
\end{eqnarray}
where
$h = 2\pi HK_{x^2}/\Phi_0$ is a dimensionless magnetic field and 
\begin{eqnarray}
k_{40} & = & 3(K_{x^4}+K_{x^2y^2})/2 K_{x^2}^2 \ , \nonumber \\  
k_{44} & = & (K_{x^4}-3K_{x^2y^2})/2 K_{x^2}^2 \ ,  \nonumber \\
k_{60} & = & (2K_{x^6}+3 K_{x^4y^2}+3 K_{x^2y^4})/8 K_{x^2}^3 \ , 
\label{knm}\\
k_{62} & = & (K_{x^4y^2} - K_{x^2y^4})/8 K_{x^2}^3\ , \ \ 
k_{66} = -15 k_{62} \ , \nonumber \\
k_{64} & = & (2K_{x^6}-5 K_{x^4y^2}-5 K_{x^2y^4})/8 K_{x^2}^3\ .
\nonumber 
\end{eqnarray}
The level-number operator $\hat{n}=\hat{a}^{\dagger}\hat{a}$ and its polynomials
$\hat{n}_{60} =( 20 \hat{n}^3 + 30 \hat{n}^2 + 40 \hat{n} + 15)$, $\hat{n}_{62}
=( 15 \hat{n}^2 + 63 \hat{n} + 45)$, $\hat{n}_{64} =( 6 \hat{n} + 15)$ are
invariant with respect to an arbitrary rotation about the $\hat{z}$-axis. 
The discrete rotations of the point crystal
group are responsible for the appearance of $\hat{a}^n$ terms.
In particular, 
the operators $\hat{a}^2$ and $\hat{a}^6$ break the four-fold rotational
symmetry about $\hat{z}$ and discriminate between the  $\hat{x}$- 
and the $\hat{y}$-axes.

We use the standard procedure to
determine the FLL geometry in the vicinity of the upper critical field
\cite{abrikosov,saint-james}. Previously,
such an approach has been applied for
superconductors with tetragonal \cite{park,klironomos,chang,agterberg},
orthorhombic \cite{han}, and hexagonal \cite{zhitomirsky} 
crystal structures. 
Solution of the linearized GL equations obtained from Eqs.~(\ref{GL})
and (\ref{F246fin})  is expanded in the Landau levels 
up to the sixth order:
\begin{equation}
\Psi = \lambda \psi \ ,\ \psi = f_0+c_2 e^{2i\varphi}f_2+c_4e^{4i\varphi}
f_4 + c_6 e^{6i\varphi} f_6 \ , 
\label{expan}
\end{equation}
where $f_{n} =  (\hat{a}^\dagger)^n f_0/\sqrt{n!}$ and $\hat{a}f_0=0$.  
In order to construct a 
periodic vortex structure at the zeroth Landau level $f_0$
it is convenient to go from a laboratory frame determined by
the crystal axes to a rotated frame such that a new $\hat{x}$-axis
points between a pair of nearest-neighbor vortices and is rotated by angle
$\varphi$ with respect to the crystal axis.
The vector potential is chosen in the Landau gauge 
${\bf A}=(-Hy,0,0)$ and the periodic solution with 
one flux quantum per unit cell is written as
\begin{equation}
f_0({\bf r})=\sum_m\exp\Bigl[- \pi i\rho m^2 + \frac{2\pi}{a}imx
-\frac{\pi H}{\Phi_0}(y-ma\sigma)^2\Bigr].
\end{equation}
The basis vectors of the FLL are $(a,0,0)$ and 
$(a\rho,a\sigma,0)$ in the rotated frame, which satisfy the 
one-flux-quantum condition
$Ha^2\sigma = \Phi_0$.

\begin{figure}[t]
\begin{center}
\includegraphics[width=0.9\columnwidth]{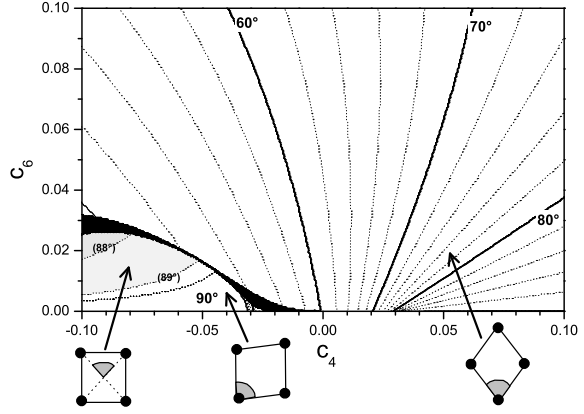}
\end{center}
\caption{Fig.1: 
Geometry of the FLL unit cell as a function of $c_4$ and $c_6$.
Lines indicate corresponding values of angles, which parameterize
a rhombic or a rectangular unit cells
}
\label{fig:phase}
\end{figure}

The expansion coefficients in Eq.~(\ref{expan}) are found
for  the eigensolution of the linearized GL equation 
as a perturbation expansion in a small
parameter $\alpha$:
\begin{equation}
c_4 =\frac{\sqrt{6}}{4} k_{44}\alpha  \ , \ 
c_2 = -\frac{45}{2\sqrt{2}}k_{62}\alpha^2 \ , \  
c_6 =  - \sqrt{5} k_{66} \alpha^2  \ , 
\end{equation}
whereas  the upper critical field is given by
\begin{equation}
h_{c2} = -\alpha - k_{40} \alpha^2 + (15 k_{60} -2 k_{40}^2 - 3
k_{44}^2) \alpha^3 \ .
\end{equation}
Neglecting the magnetic field
contribution to the free energy $h^2/8\pi$ in the 
large-$\kappa$ limit we obtain for the energy density
\begin{equation}
\langle F_{GL} \rangle = \lambda^2 
\langle\psi^*(\alpha+{\cal F}_{\Pi})\psi
\rangle 
 + \lambda^4
\frac{\beta}{2}  \langle |\psi|^4 \rangle \ ,
\label{eq:energy} 
\end{equation}
where $\langle f \rangle = (1/V)\int dr^3 f({\bf r})$. The quadratic
term in the above equation is calculated as 
$\langle|f_0|^2\rangle(h-h_{c2})(1 -2k_{40}\alpha)$.  
Then, straightforward  minimization of the  energy
density (\ref{eq:energy}) with respect to $\lambda$ yields 
\begin{equation}
\langle F_{GL} \rangle = - \frac{(h-h_{c2})^2 (1-2k_{40}
\alpha)^2} {2\beta\langle|\psi|^4 \rangle/\langle|f_0|^2
\rangle^2 } \ .
\end{equation}
The equilibrium distribution of the order parameter $\Psi({\bf r})$
is found by  
minimizing the geometrical factor 
\begin{equation} 
\beta_A = \langle |\psi|^4\rangle/\langle|f_0|^2\rangle^2 \ .
\end{equation}
This generalized Abrikosov's parameter is a function of only three 
variables $\rho$, $\sigma$,
and $\varphi$. An explicit calculation yields
\begin{equation}
\beta_A = \sqrt{\sigma} \sum_{m,n } \exp{\left[
2\pi i\rho(m^2\!-n^2) - 2\pi\sigma(m^2\!+n^2) \right]}\, I_{m,n},
\label{betaA}
\end{equation}
where summation goes over all integer and half-integer pairs
$(m,n)$.
Function $I_{m,n}$ is defined by an integral 
\begin{eqnarray}
I_{m,n} &=& \sqrt{\frac{2}{\pi}}
\int dye^{-2y^2}P\bigl(y+\sqrt{2\pi\sigma}m\bigr)\,
P\bigl(y-\sqrt{2\pi\sigma}m\bigr) \nonumber \\
&&\mbox{}\times P^*(y+\sqrt{2\pi\sigma}n)
P^*(y-\sqrt{2\pi\sigma}n) \ ,
\end{eqnarray} 
where
\begin{equation}
P(y)= 1 + c_2 e^{2i\varphi} H_2(y) + c_4 e^{4i\varphi} H_4(y) + c_6
e^{6i\varphi} H_6(y)
\end{equation}
and $H_n(y)$ are the Hermite's polynomials.
If we keep, for simplicity, only the terms, which are linear in
$c_n$, then the Abrikosov parameter is reduced to
\begin{equation}
\beta_A \approx \beta_0 + 4 {\Real} ( c_2 e^{2i\varphi} \beta_2 + c_4
e^{4i\varphi} \beta_4 + c_6 e^{6i\varphi} \beta_6) \ ,
\end{equation} 
where 
\begin{equation}
\beta_k = \langle f_0^{*2} f_0 f_k\rangle/\langle|f_0|^2\rangle^2 \ .
\label{betas}
\end{equation}
Function $\beta_0(\rho,\sigma)$ is the standard energy parameter
for an isotropic superconductor \cite{abrikosov}, which 
is given by Eq.~(\ref{betaA}) with $I^{(0)}_{m,n}=1$ \cite{saint-james}. 
The other functions $\beta_k$ are obtained from Eq.~(\ref{betaA})
by substituting the corresponding $I^{(k)}_{m,n}$:
\begin{eqnarray}
I^{(2)}_{m,n} &=&  \frac{1}{\sqrt{2}} \Bigl(4\pi\sigma n^2-\frac{1}{2}\Bigr)\ ,
\nonumber \\
I^{(4)}_{m,n} &=& \frac{1}{\sqrt{6}} \Bigl(
8\pi^2\sigma^2n^4 - 6\pi\sigma n^2 + \frac{3}{8}\Bigr) \  ,  \\
I^{(6)}_{m,n} &=& \frac{1}{3\sqrt{5}} \Bigl(
16\pi^3\sigma^3n^6 - 30 \pi^2\sigma^2n^4
+ \frac{45}{4} \pi \sigma n^2 - \frac{15}{32}  \Bigr).
\nonumber
\end{eqnarray}
Hexagonal vortex lattice with $\rho=1/2$ and $\sigma=\sqrt{3}/2$
corresponds to the absolute minima of 
functions $\beta_0(\rho,\sigma)$ and $\beta_6(\rho,\sigma)$.
The other two functions $\beta_4$ and $\beta_2$ tend to stabilize 
a square and a distorted triangular lattices, respectively.
Their competition yields a rich phase diagram
of a tetrahedral superconductor.
In finding the lowest energy vortex lattice
we have kept also additional terms in $\beta_A$, which are
proportional to high-orders of $c_n$.

\begin{figure}[tbp]
\begin{center}
\includegraphics[width=0.9\columnwidth]{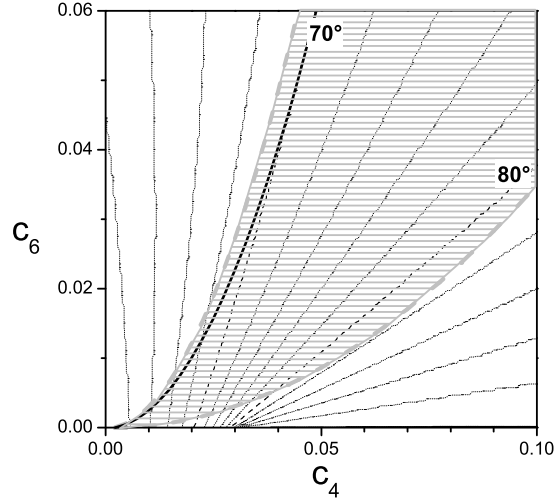}
\end{center}
\caption{Fig.2: Expansion of the right part of fig.1.
The shaded region is composed 
by parabolas $c_6(T) \propto c_4(T)^2$;
the black line is defined by $c_6=25c_4^2$  }
\label{fig:stable}
\end{figure} 

Equilibrium form and orientation of the vortex
lattice in a tetrahedral superconductor should be obtained
by minimizing $\beta_A(\rho,\sigma,\varphi)$ for different sets 
of the parameters $\alpha$ and $k_{ij}$ in the GL functional.
The task is considerably simplified if the expansion of $\psi$ 
in Eq.~(\ref{expan}) is rapidly converging, {\it e.g.},
near $T_c$ or for a superconductor with a weak anisotropy. 
In such a case, a set of the effective parameters
can be reduced to $c_4$, which is determined by the lowest order
anisotropy in quartic terms, and  $c_6$, which quantifies the $x$--$y$
discrimination introduced by the tetrahedral symmetry.  
Since $c_2\propto c_6$ in the leading order in $\alpha$, we fix $c_2 =
-3/(2\sqrt{10}) c_6\approx -0.47 c_6$.
The energy parameter $\beta_A$ is, then, numerically minimized with the
conjugate gradient method. The parameter space is further restricted to
$c_6 \geq 0$ since a change of the sign of $c_6$ corresponds to 
a rotation of the coordinate frame by
$\varphi=90^{\circ}$. 

Our main results are presented in fig.\ref{fig:phase}, where
the geometry adopted by the FLL is plotted for different values of $c_4$ and
$c_6$. The considered region in the parameter space 
is divided into two parts corresponding to
highly symmetric vortex lattices separated by a transition
region colored in black. 
In the main area, the whole $(c_4,c_6)$ plane  apart from the lower
left corner, the unit cell has a rhombic
shape with the longest diagonal parallel to the $\hat{y}$-axis.
The apex angle of a rhombus $\theta$ varies from 50$^{\circ}$ to 90$^{\circ}$
as indicated by thin dotted lines.

Since $c_4={\cal O}(\alpha)$ and $c_6={\cal O} (\alpha^2)$, an evolution 
along the $H_{c2}(T)$-curve follows approximately a parabolic path 
in the $(c_4,c_6)$-phase diagram, which starts at the origin
$c_4=0$ and $c_6=0$. For a certain range of 
parameters, the apex angle $\theta$  along such an $H_{c2}(T)$-path
does not change  significantly with temperature. 
For example, $\theta$ remains between $70^{\circ}$ and $80^{\circ}$ in the
shaded area of  fig.\ref{fig:stable} composed by a set of parabolas 
given by $c_6/c_4^2 =-16\sqrt{5}k_{66}/6k_{44}^2$  with
\begin{equation}
-5 \leq k_{66}/k_{44}^2 \leq -0.6  \ . 
\end{equation}
The reason for such a weak dependence is a cancellation of two
opposite tendencies determined by $\beta_2(\rho,\sigma)$ and 
$\beta_4(\rho,\sigma)$, see Eq.~(\ref{betas}).
If the crystal would have the cubic or the 
tetragonal symmetry, $c_6$ is zero and the apex angle increases almost
linearly between $60^{\circ}$ and $90^{\circ}$, with a phase transition
into a square lattice afterwards.
The present investigation shows that the tetrahedral symmetry 
introduces a fundamental difference via a sixth-order gradient terms.

\begin{figure}[b]
\begin{center}
\includegraphics[width=0.9\columnwidth]{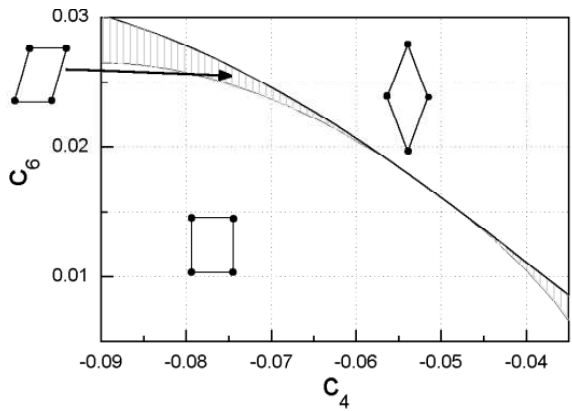}
\end{center}
\caption{Fig.3: 
Expansion of the left side of fig.1: transition
region  between the vortex unit cells of rhombic and rectangular shapes
}
\label{fig:transition}
\end{figure} 
 
In the lower left corner of the phase diagram, 
for $c_4<0$ and small $c_6$, the shape of the vortex unit cell
changes via a second-order transition from a rhombus
with the longest diagonal oriented 
by $45^{\circ}$ from the $\hat{x}$-axis  to a rectangle 
with the longest side along the $\hat{y}$-axis. When 
$c_4$ decreases, 
the small apex angle of the rhombus goes from 60$^{\circ}$ to
90$^{\circ}$. The two regions on the phase diagram
are separated by a transition region (shown in black), where the unit
cell of FLL has no mirror symmetry.
Fig.\ref{fig:transition} illustrates that a
transformation of the unit cell corresponds to a shear instability
of the vortex lattice: vortex chains in the rectangular lattice slide
along one of the directions such that the lattice 
turns into a centered rectangular (rhombic) lattice with the same
volume of the unit cell 
and the same side ratio. This transformation goes via two 
second-order transitions
for $c_4<-0.06$, whereas for 
$-0.055<c_4< -0.048$ the transition is of the first order.
As a last remark, we would like to point at a peculiar possibility
for a tetrahedral superconductor with
negative $c_4$. While moving along the $H_{c2}(T)$ line towards
low temperatures,  the vortex lattice first 
changes from a triangular one
into a rectangular lattice and then transforms back into a distorted
triangular lattice.

Finally, we shall apply the above results to PrOs$_4$Sb$_{12}$. 
Topology of the Fermi 
surface of this material has been studied by the de Haas-van Alphen 
measurements and compared to the results of the LDA+$U$ band structure
calculations by Sugawara {\it et al.} \cite{sugawara,harima}.
The Fermi surface is composed of three sheets: one of the 48th band
and two of the 49th band. The contribution to the density of states 
from the 48th band is relatively small: $N_{48}(0)\sim 0.04N_{49}(0)$.
Besides, only a small Fermi velocity in the 49th band is capable to 
account for a large value of the upper critical field at zero temperature
$H_{c2}(0) = 2.2$~T \cite{bauer}.
Accordingly, we assume that the active band, which is responsible for
superconductivity in PrOs$_4$Sb$_{12}$, is the 49th band, whereas 
the 48th band plays a passive role and may have a smaller
gap amplitude as suggested by Seyfarth {\it et al.} \cite{seyfarth}.

\begin{table}
\caption{Table I: Fermi surface averages of different combinations of the 
components of the Fermi velocity 
$\langle v_{Fx}^kv_{Fy}^lv_{Fz}^m\rangle_{\rm FS}$ for the 49th band 
in units of ($10^5$~m/s)$^{k+l+m}$.}
\begin{tabular}{rc|cccccc}
$klm$ &\ & \ & $\alpha$-sheet &\ \ & $\gamma$-sheet &\ \ &  combined  \\[1mm]
\hline
&&&&&&&\\[-4mm]
200 & & & $2.19$  & & $1.46$ & & $1.61$ \\
400 & & & $14.7$ & & $6.36$ & & $8.11$ 
\\
220 & & & $1.51$  & & $1.63$ & & $1.60$ \\
600 & & & $127.$ & & $45.0$ & & $62.2$
\\
420 & & & $5.69$  & & $3.74$ & & $4.15$ \\
240 & & & $4.49$  & & $8.75$ & & $7.85$ \\
222 & & & $1.03$  & & $0.63$ & & $0.714$ 
\end{tabular}
\end{table}

The band structure data for 195 $\bf k$-points in the irreducible part
of the Brillouin zone of PrOs$_4$Sb$_{12}$ have been previously
used in the comparison between LDA results and de Haas-van Alphen
data \cite{sugawara}. We interpolated 
between these data points using appropriate lattice harmonics 
and, then, calculated 
numerically the Fermi surface averages for the 49th band
using a much smaller mesh in the momentum space.
The combined averages have been obtained by a sum of the
two contributions weighted according to partial densities
of states $N^\alpha_{49}/N^\gamma_{49}\approx 0.27$. The results are
summarized in Table I. Using Eqs.~(\ref{Kij}) and (\ref{knm}) 
we find for the dimensionless gradient constants $k_{44}=-0.29$, 
$k_{66}=-0.35$ such that $k_{66}/k_{44}^2=-4.3$ and
\begin{equation}
c_6(T)/c_4(T)^2 \approx 25.0 \ .
\end{equation}
The above relation is illustrated in fig.\ref{fig:stable} by a solid line.
For $c_4(T)\ge 0.02$, which corresponds to $|\alpha|\ge 0.12$ and 
$T<0.9T_c$, the vortex lattice has a rhombic unit cell with a nearly
$T$-independent apex angle $\theta=70^{\circ}$.  For 
a rhombic lattice this angle coincides exactly with the angle
between two reciprocal lattice vectors. 
The latter angle has been measured in
the neutron scattering experiment \cite{huxley} and is equal to
$\theta\approx 75\pm 5^\circ$.
Thus, there is a good agreement between our calculation and 
the experimental data on PrOs$_4$Sb$_{12}$.
We emphasize again, that in a tetrahedral superconductor such a stable
distortion of a hexagonal vortex lattice arises due to a competition
between several anisotropic gradient terms.
When comparing  the above results with experimental data one should,
of course, bear in mind that applicability regions are somewhat different.
Formally, our calculation has been restricted
to the vicinity of $H_{c2}$, but is valid in a wider range of applied
fields $H>0.2H_{c2}$ in high-$\kappa$ materials such as PrOs$_4$Sb$_{12}$ 
($\kappa \approx 29$). The temperature range is in the GL regime
$0.6T_c<T<0.9T_c$, whereas the neutron measurements have been performed
for  $T<0.45T_c$. Further investigations of the shape of the vortex
lattice in tetrahedral superconductors 
in the London limit at low temperatures would be, therefore,
useful. 

A different  confirmation of our analysis can be provided 
by angle dependence of the upper critical field since 
the amplitude of the modulations are related to the anisotropy of the Fermi
surface \cite{takanaka,teichler,dao}.
For example, a perturbation calculation
yields the main contribution to the [001]-plane four-fold modulation:
\begin{equation}
\frac{H_{c2}(\phi)}{H_{c2}(0)} = 1 +
\frac{3}{8}\,\alpha k_{44} \bigl(\cos{4\phi}-1\bigr) \ ,
\end{equation}
where $\phi$ is angle between an applied magnetic 
field and the $\hat{x}$-axis. 

We are indebted to V. P. Mineev for stimulating discussion 
and to H. Harima for providing us the band structure data for 
PrOs$_4$Sb$_{12}$.

\end{document}